\documentclass{article}
\usepackage{graphicx}
\usepackage[centertags]{amsmath}
\usepackage{amsfonts}
\usepackage{amssymb}
\usepackage{amsthm}

\title{Quasiclassical and Quantum Systems of Angular Momentum. Part II. Quantum Mechanics on Lie Groups and Methods of Group Algebras}

\author{J. J. S\l awianowski, V. Kovalchuk, A. Martens,\\ 
B. Go\l ubowska, and E. E. Ro\.zko\\
Institute of Fundamental Technological Research,\\
Polish Academy of Sciences,\\
$5^{\rm B}$, Pawi\'{n}skiego str., 02-106 Warsaw, Poland\\
e-mails: jslawian@ippt.gov.pl, vkoval@ippt.gov.pl,\\ 
amartens@ippt.gov.pl, bgolub@ippt.gov.pl, erozko@ippt.gov.pl}

\begin{document}

\maketitle
\begin{abstract}
In Part I of this series we presented the general ideas of applying group-algebraic methods for describing quantum systems. The treatment was there very "ascetic" in that only the structure of a locally compact topological group was used. Below we explicitly make use of the Lie group structure. Basing on differential geometry enables one to introduce explicitly representation of important physical quantities and formulate the general ideas of quasiclassical representation and classical analogy.
\end{abstract}

Let us now discuss the very important situation when $G$ is a compact
Lie group. The special stress is laid on semisimple Lie groups or
their central extension. We are particularly interested in problems
concerning angular momentum, i.e., the group ${\rm SU}(2)$ or its quotient
${\rm SO}\left(3,\mathbb{R}\right)={\rm SU}\left(2\right)/Z_{2}$. Nevertheless,
it is convenient to begin with remarks concerning the general situation.

The Lie group of $G$ will be denoted by $G'$. We assume $G$ to
be a linear group, i.e., a group of finite matrices, some subgroup
of ${\rm GL}\left(N,\mathbb{R}\right)$ or ${\rm GL}\left(N,\mathbb{C}\right)$.
This simplifies notation. Of curse, any compact Lie group is linear.
Lie algebras are meant in the matrix commutator sense. Let $\left(\ldots,e_{a},\ldots\right)$
be some basis in $G'$, the structure constants are meant in the following convention:
\begin{equation}
\left[e_{a},e_{b}\right]=e_{a}e_{b}-e_{b}e_{a}=e_{k}C^{k}{}_{ab}.\label{eq_2.1}
\end{equation}
The Killing metric tensor on $G'$, i.e., the ${\rm Ad}$-invariant scalar
product $\gamma$, is meant in the following convention: 
\begin{equation}
\gamma\left(u,v\right)={\rm Tr}\left({\rm ad}_{u}{\rm ad}_{v}\right),\label{eq_2.2}
\end{equation}
where ${\rm ad}_{u}\in L\left(G'\right)\simeq G'\otimes G^{\prime\ast}$ is
given by the usual formula: 
\begin{equation}
{\rm ad}_{u}\cdot x=\left[u,x\right].\label{eq_2.3}
\end{equation}
Analytically, in terms of the basis $e$, 
\begin{equation}
\gamma_{ab}=C^{k}{}_{la}C^{l}{}_{kb},\qquad \gamma=\gamma_{ab}e^{a}\otimes e^{b},\label{eq_2.4}
\end{equation}
where $e^{a}\in G^{\prime\ast}$ are elements of the dual basis,
$\left\langle e^{a},e_{b}\right\rangle =\delta^{a}{}_{b}$. If $G$
is compact and semisimple, then $\gamma$ is negatively definite and in an
appropriate basis $e$, $\gamma_{ab}$ is a negative multiple of $\delta_{ab}$.
Usually the basis is chosen in some convenient way motivated by various reasons, then it is customary to change the normalization
of $\gamma_{ab}$ replacing it just by $g_{ab}=\delta_{ab}$.
The contravariant inverse of $\gamma$, $\gamma^{-1}\in G'\otimes G'$,
is analytically given by 
\begin{equation}
\gamma^{-1}=\gamma^{ab}e_{a}\otimes e_{b},\qquad \gamma^{ac}\gamma_{cb}=\delta^{a}{}_{b}.\label{eq_2.5}
\end{equation}
 
In the trivial central extension $G\times U\left(1\right)$ of $G$,
the Killing tensor is degenerate and $U\left(1\right)'$ is the degenerate
direction of $\left(G\times U\left(1\right)\right)'$. Then it is
customary to use the metric tensor obtained as a direct combination
of the Killing metric on $G'$ and the invariant metric on $U\left(1\right)'$;
the latter is unique up to normalization. Sometimes one proceeds similarly when dealing with direct or semidirect products of semisimple groups and Abelian ones of arbitrary dimension, however, if that dimension is higher than one, the
Abelian component of metric has a non-canonical arbitrariness. 

Canonical coordinates of the first kind $k^{a}$ are defined by the formula:
\begin{equation}
g\left(k^{1},\ldots,k^{n}\right)=\exp\left(k^{a}e_{a}\right),\qquad \dim G=n,\label{eq_2.6}
\end{equation}
obviously, the summation convention is used on the right-hand side.
This choice is often convenient, but also other ones are useful, e.g.,
canonical coordinates of the second kind, 
\begin{equation}
g\left[\xi^{1},\ldots,\xi^{n}\right]=
\exp\left(\xi^{1}e_{1}\right)\ldots\exp\left(\xi^{n}e_{n}\right),\label{eq_2.7}
\end{equation}
or something between, like Euler angles on ${\rm SO}\left(3,\mathbb{R}\right)$
or ${\rm SU}\left(2\right)$. Often some generalized coordinates, "curvilinear"
with respect to $k^{a}$ or $\xi^{a}$, are better suited to particular
problems. In any case, the choice of coordinates is a matter of convenience.

The differential structure of $G$ offers some powerful tools of analysis.
First of all, one uses differential operators generating transformations
(117), (118) in \cite{1}. Generators of left and right regular translations
are defined in the convention:
\begin{eqnarray}
\left(\mathcal{L}_{a}\psi\right)\left(g\left(\overline{k}\right)\right)&=& \frac{\partial}{\partial x^{a}}\left(\psi\left(g\left(\overline{x}\right)g\left(\overline{k}\right)\right)
\right)\bigr|_{\overline{x}=0},\label{eq_2.8a}\\
\left(\mathcal{R}_{a}\psi\right)\left(g\left(\overline{k}\right)\right)&=& \frac{\partial}{\partial x^{a}}\left(\psi\left(g\left(\overline{k}\right)g\left(\overline{x}\right)
\right)\right)\bigr|_{\overline{x}=0},\label{eq_2.8b}
\end{eqnarray}
i.e., roughly, we have the following expansions for small values
of the group parameters $\overline{\varepsilon}$:
\begin{eqnarray}
\psi\left(g\left(\overline{\varepsilon}\right)g\right)&\approx& \psi\left(g\right)+\varepsilon^{a}\left(\mathcal{L}_{a}\psi\right)\left(g\right),\label{eq_2.9a}\\
\psi\left(gg\left(\overline{\varepsilon}\right)\right)&\approx& \psi\left(g\right)+\varepsilon^{a}\left(\mathcal{R}_{a}\psi\right)\left(g\right),\label{eq_2.9b}
\end{eqnarray}
valid to terms quadratic and higher order in $\overline{\varepsilon}$. 

$\mathcal{L}_{a}$, $\mathcal{R}_{a}$ are respectively, basic right- and left-invariant vector fields on $G$. We represent them as follows: \begin{equation}
\mathcal{L}_{a}=\mathcal{L}^{i}{}_{a}\left(\overline{k}\right)
\frac{\partial}{\partial k^{i}},\qquad \mathcal{R}_{a}=\mathcal{R}^{i}{}_{a}
\left(\overline{k}\right)\frac{\partial}{\partial k^{i}}.\label{eq_2.10}
\end{equation}
With this convention we have the following commutation rules: 
\begin{equation}
\left[\mathcal{L}_{a},\mathcal{L}_{b}\right]=-C^{k}{}_{ab}\mathcal{L}_{k},\qquad
\left[\mathcal{R}_{a},\mathcal{R}_{b}\right]=C^{k}{}_{ab}\mathcal{R}_{k},\qquad
\left[\mathcal{L}_{a},\mathcal{R}_{b}\right]=0.\label{eq_2.11}
\end{equation}
Similarly one defines differential operators $\mathcal{D}_{a}$ generating
inner automorphisms,
\begin{equation}
\left(\mathcal{A}_{a}\psi\right)\left(g\left(\overline{k}\right)\right)=
\frac{\partial}{\partial x^{a}}\left(\psi\left(g\left(\overline{x}\right)g\left(\overline{k}\right)g
\left(-\overline{x}\right)\right)\right)\Bigr|_{\overline{x}=0},\label{eq_2.12}
\end{equation}
i.e., roughly,
\begin{equation}
\psi\left(g(\overline{\varepsilon})gg(-\overline{\varepsilon})\right)\approx
\psi(g)+\varepsilon^{a}\left(\mathcal{A}_{a}\psi\right)(g),\label{eq_2.13}
\end{equation}
up to higher-order corrections in $\overline{\varepsilon}$. Obviously,
\begin{equation}
\mathcal{A}_{a}=\mathcal{L}_{a}-\mathcal{R}_{a},\label{eq_2.14}
\end{equation}
and we use the notation 
\begin{equation}
\mathcal{A}_{a}=\mathcal{A}^{i}{}_{a}(\overline{k})
\frac{\partial}{\partial k^{i}}.\label{eq_2.15}
\end{equation}
It is also clear that 
\begin{equation}
\left[\mathcal{A}_{a},\mathcal{A}_{b}\right]=-C^{k}{}_{ab}\mathcal{A}_{k}
\label{eq_2.16}
\end{equation}
and 
\begin{equation}
\left[\mathcal{A}_{a},\mathcal{L}_{b}\right]=-C^{k}{}_{ab}\mathcal{L}_{k},\qquad
\left[\mathcal{A}_{a},\mathcal{R}_{b}\right]=C^{k}{}_{ab}\mathcal{R}_{k}.
\label{eq_2.17}
\end{equation}
The $\pm$ signs on the right-hand sides of (\ref{eq_2.11}), (\ref{eq_2.16}),
(\ref{eq_2.17}) are essential. As mentioned, the translation operators
(117), (118), (150), (151) in \cite{1} are unitary in $L^{2}(G)$
due to the very definition and properties of the Haar measure on $G$.
Therefore, their generators $\mathcal{L}_{a}$, $\mathcal{R}_{a}$, $\mathcal{A}_{a}$ are skew-symmetric in the corresponding dense subdomain of $L^{2}(G)$,
\begin{eqnarray}
\left\langle \mathcal{L}_{a}\psi|\varphi\right\rangle  & = & 
-\left\langle \psi|\mathcal{L}_{a}\varphi\right\rangle,\label{eq_2.18a}\\
\left\langle \mathcal{R}_{a}\psi|\varphi\right\rangle  & = & 
-\left\langle \psi|\mathcal{R}_{a}\varphi\right\rangle,\label{eq_2.18b}\\
\left\langle \mathcal{A}_{a}\psi|\varphi\right\rangle  & = & 
-\left\langle \psi|\mathcal{A}_{a}\varphi\right\rangle;\label{eq_2.18c} 
\end{eqnarray}
by their very definition as differential operators, $\mathcal{L}_{a}$, $\mathcal{R}_{a}$, $\mathcal{A}_{a}$ are not globally defined on $L^{2}(G)$. 

Let us quote the following formulas:
\begin{eqnarray}
L\left[g(\overline{k})^{-1}\right]  =L\left[g(-\overline{k})\right]&=& \exp\left(k^{a}\mathcal{L}_{a}\right),\label{eq_2.19a}\\
R\left[g(\overline{k})^{-1}\right]  =R\left[g(-\overline{k})\right]&=& \exp\left(k^{a}\mathcal{R}_{a}\right),\label{eq_2.19b}\\
A\left[g(\overline{k})^{-1}\right]  =A\left[g(-\overline{k})\right]&=& \exp\left(k^{a}\mathcal{A}_{a}\right),\label{eq_2.19c}
\end{eqnarray}
which hold when their right-hand sides are well defined; thus, in an appropriate
dense subdomain, with the convergence meant in the sense of $L^{2}(G)$-norm. Obviously, the left-hand sides are well-defined in action on
the total linear space of all possible functions on $G$ (with arbitrary
target spaces, not necessarily $\mathbb{C}$).

The imaginary-unit multiples of $\mathcal{L}_{a}$, $\mathcal{R}_{a}$, $\mathcal{A}_{a}$ are formally Hermitian (symmetric). Because of the obvious physical reasons we introduce the formally Hermitian operators of $L[G]$-, $R[G]$- and $A[G]$-momenta, just the quantum versions of
the corresponding classical momentum mappings, 
\begin{equation}
\mathbf{\Sigma}_{a}=\frac{\hbar}{i}\mathbf{\mathcal{L}}_{a,}\qquad
\mathbf{\widehat{\Sigma}}_{a}=
\frac{\hbar}{i}\mathbf{\mathcal{R}}_{a,}\qquad
\mathbf{\Delta}_{a}=\frac{\hbar}{i}\mathbf{\mathcal{A}}_{a}=
\mathbf{\Sigma}_{a}-\mathbf{\widehat{\Sigma}}_{a}.\label{eq_2.20}
\end{equation}
As operators acting on $L^{2}(G)$-wave functions, they satisfy the
obvious quantum Poisson brackets:
\begin{eqnarray}
\frac{1}{\hbar i}\left[\mathbf{\Sigma}_{a},\mathbf{\Sigma}_{b}\right]&=& \left\{ \mathbf{\Sigma}_{a},\mathbf{\Sigma}_{b}\right\} _{Q}= C^{k}{}_{ab}\mathbf{\Sigma}_{k},\label{eq_2.21a}\\
\frac{1}{\hbar i}\left[\mathbf{\widehat{\Sigma}}_{a},\mathbf{\widehat{\Sigma}}_{b}\right]&=& \left\{ \mathbf{\widehat{\Sigma}}_{a},\mathbf{\widehat{\Sigma}}_{b}\right\} _{Q}= -C^{k}{}_{ab}\mathbf{\widehat{\Sigma}}_{k},\label{eq_2.21b}\\
\frac{1}{\hbar i}\left[\mathbf{\Sigma}_{a},\mathbf{\widehat{\Sigma}}_{b}\right]&=& 0.\label{eq_2.21c}
\end{eqnarray}
The corresponding classical counterparts are given by the phase-space
functions:
\begin{equation}
\Sigma_{a}=p_{i}\mathcal{L}^{i}{}_{a},\qquad \widehat{\Sigma}_{a}=
p_{i}\mathcal{R}^{i}{}_{a},\qquad \Delta_{a}=p_{i}\Delta^{i}{}_{a}=
\Sigma_{a}-\widehat{\Sigma}_{a}.\label{eq_2.22}
\end{equation}
Their classical Poisson brackets are structurally identical with (\ref{eq_2.21a})--(\ref{eq_2.21c}), i.e.,
\begin{equation}
\left\{ \Sigma_{a},\Sigma_{b}\right\} =C^{k}{}_{ab}\Sigma_{k},\qquad 
\left\{ \widehat{\Sigma}_{a},\widehat{\Sigma}_{b}\right\} =-C^{k}{}_{ab}\widehat{\Sigma}_{k},\qquad 
\left\{ \Sigma_{a},\widehat{\Sigma}_{b}\right\}=0.\label{eq_2.23}
\end{equation}
As mentioned, the regular translations and automorphisms (117), (118) in \cite{1},
and all operators of convolution (127), (132), (143),
(145) in \cite{1} preserve separately all subspaces/minimal ideals
$M(\alpha)$. This is also true for the operators $\mathcal{L}_{a}$, $\mathcal{R}_{a}$, $\Delta_{a}$
as generators of those group actions. Obviously, their multiples
$\Sigma_{a}$, $\widehat{\Sigma}_{a}$, $\Delta_{a}$ also preserve all ideals
$M(\alpha)$. The basic right- and left-invariant differential forms
on $G$ will be denoted respectively by $\mathcal{L}^{a}$, $\mathcal{R}^{a}$;
by definition they are assumed to be dual to $\mathcal{L}_{a}$, $\mathcal{R}_{a}$:
\begin{equation}
\left\langle \mathcal{L}^{a},\mathcal{L}_{b}\right\rangle =\left\langle \mathcal{R}^{a},\mathcal{R}_{b}\right\rangle =\delta^{a}{}_{b}.\label{eq_2.24}
\end{equation}
We shall use the standard analytical representation dual to (\ref{eq_2.10}),
\begin{equation}
\mathcal{L}^{a}=\mathcal{L}^{a}{}_{i}(\overline{k})dk^{i},\qquad
\mathcal{R}^{a}=\mathcal{R}^{a}{}_{i}(\overline{k})dk^{i},\label{eq_2.25}
\end{equation}
where obviously 
\begin{equation}
\mathcal{L}^{i}{}_{a}\mathcal{L}^{a}{}_{j}=
\mathcal{R}^{i}{}_{a}\mathcal{R}^{a}{}_{j}=\delta^{i}{}_{j},\qquad
\mathcal{L}^{a}{}_{i}\mathcal{L}^{i}{}_{b}=
\mathcal{R}^{a}{}_{i}\mathcal{R}^{i}{}_{b}=\delta^{a}{}_{b}.\label{eq_2.26}
\end{equation}
Obviously, the following equations are satisfied, dual to (\ref{eq_2.11}):
\begin{equation}
d\mathcal{L}^{a}=\frac{1}{2}C^{a}{}_{bd}\mathcal{L}^{b}\wedge\mathcal{L}^{d},\qquad d\mathcal{R}^{a}=-\frac{1}{2}C^{a}{}_{bd}\mathcal{R}^{b}\wedge\mathcal{R}^{d}.
\label{eq_2.27}
\end{equation}
Let us notice that 
\begin{equation}
\mathcal{L}_{a}(g)=\left({\rm Ad}_{g^{-1}}\right)^{b}{}_{a}\mathcal{R}_{b}(g),
\label{eq_2.28}
\end{equation}
where, obviously, the matrices $\left[\left({\rm Ad}_{g}\right)^{b}{}_{a}\right]$ are implicitly given by: 
\begin{equation}
{\rm Ad}_{g}e_{a}=ge_{a}g^{-1}=e_{b}\left({\rm Ad}_{g}\right)^{b}{}_{a}.\label{eq_2.29}
\end{equation}
Similarly, $\left({\rm ad}_{y}\right)^{b}{}_{a}$ are given by 
\begin{equation}
{\rm ad}_{y}e_{a}=\left[y,e_{a}\right]=e_{b}\left({\rm ad}_{y}\right)^{b}{}_{a},
\label{eq_2.30}
\end{equation}
thus, 
\begin{equation}
\left({\rm ad}_{y}\right)^{b}{}_{a}=C^{b}{}_{da}y^{d}=-C^{b}{}_{ad}y^{d},\label{eq_2.31}
\end{equation}
and 
\begin{equation}
{\rm Ad}_{\exp(a)}=\exp\left({\rm ad}_{a}\right);\label{eq_2.32}
\end{equation}
in finite dimensions all above expressions are well-defined. Dually
to (\ref{eq_2.28}) we have 
\begin{equation}
\mathcal{L}^{a}(g)=\left({\rm Ad}_{g}\right)^{a}{}_{b}\mathcal{R}^{b}(g).
\label{eq_2.33}
\end{equation}
The above differential operators and differential forms are a very
useful tool of analysis. When constructing important tensor fields
and differential operators on $G$ we need certain intrinsically constructed
tensors on its Lie algebra $G'$. We mean some tensors built of the
structure constants $C^{i}{}_{jk}$ with the use of universal algebraic
operations. The first of them is $C$ itself, it is a mixed tensor
once contravariant and twice covariant, $C\in G'\otimes G^{\prime\ast}\otimes G^{\prime\ast}$,
skew-symmetric in its lower indices. The next one is the Killing tensor
$\gamma\in G^{\prime\ast}\otimes G^{\prime\ast}$ (\ref{eq_2.2}), (\ref{eq_2.4})
and its inverse tensor $\gamma^{-1}\in G'\otimes G'$. One can also
construct the higher-order covariant tensors like 
\begin{equation}
\gamma(3)_{ijk}=C^{a}{}_{bi}C^{b}{}_{cj}C^{c}{}_{ak},\label{eq_2.34}
\end{equation}
and so on, e.g., 
\begin{equation}
\gamma(m)_{i_{1}\cdots i_{m}}=C^{a}{}_{bi_{1}}C^{b}{}_{ci_{2}}\ldots C^{k}{}_{li_{m-1}}C^{l}{}_{ai_{m}},\label{eq_2.35}
\end{equation}
all of them covariant and in general non-symmetric (unlike the
Killing tensor $\gamma(2)_{ij}=\gamma_{ij}$). Let us also mention
other tensors like, 
\begin{equation}
\gamma(1)_{i}=C^{a}{}_{ai},\qquad\Gamma_{ij}=C(1)_{k}C^{k}{}_{ij}=-\Gamma_{ji}.
\label{eq_2.36}
\end{equation}
If $G$ is semisimple, then the inverse tensor (\ref{eq_2.5}) does
exist and one can construct the whole ZOO of $\gamma$-tensors by the
Killing-shift of indices. And similarly when $G$ is a trivial central
extension of some semisimple group; the invariant metric tensor on
the centre is unique up to normalization.

The Killing metric tensor on $G$ is given by 
\begin{equation}
g=\gamma_{ab}\mathcal{L}^{a}\otimes\mathcal{L}^{b}=
\gamma_{ab}\mathcal{R}^{a}\otimes\mathcal{R}^{b},\label{eq_2.37}
\end{equation}
i.e., analytically,
\begin{equation}
g_{ij}=\gamma_{ab}\mathcal{L}^{a}{}_{i}\mathcal{L}^{b}{}_{j}=
\gamma_{ab}\mathcal{R}^{a}{}_{i}\mathcal{R}^{b}{}_{j}.\label{eq_2.38}
\end{equation}
It is invariant under right and left regular translations on $G$.
Usually one changes its normalization in such a way that in certain
practically useful coordinates, at the group identity $g_{ij}$ coincides
with the Kronecker $\delta_{ij}$. In particular, if $G$ is compact,
then $\gamma$, $g$ are negatively definite; it is the natural to inverse
their signs.

The most general right-invariant metric on $G$ is given by 
\begin{equation}
{}_{r}g=\varkappa_{ab}\mathcal{L}^{a}\otimes\mathcal{L}^{b},\label{eq_2.39}
\end{equation}
where the matrix $\left[\varkappa_{ab}\right]$ is non-degenerate and constant. Similarly, for the left-invariant metrics we have 
\begin{equation}
{}_{l}g=\varkappa_{ab}\mathcal{R}^{a}\otimes\mathcal{R}^{b}.\label{eq_2.40}
\end{equation}
They become identical and doubly-invariant when $\varkappa_{ab}=\gamma_{ab}$.
Obviously, the corresponding inverse contravariant metrics are given by 
\begin{eqnarray}
g^{-1}&=&\gamma^{ab}\mathcal{L}_{a}\otimes\mathcal{L}_{b}=
\gamma^{ab}\mathcal{R}_{a}\otimes\mathcal{R}_{b},\label{eq_2.41a}\\
g^{ij}&=&\gamma^{ab}\mathcal{L}^{i}{}_{a}\mathcal{L}^{j}{}_{b}=
\gamma^{ab}\mathcal{R}^{i}{}_{a}\mathcal{R}^{j}{}_{b},\label{eq_2.41b}
\end{eqnarray}
and similarly for the inverses of (\ref{eq_2.39}), (\ref{eq_2.40}),
\begin{equation}
{}_{r}g^{-1}=\varkappa^{-1ab}\mathcal{L}_{a}\otimes\mathcal{L}_{b},\qquad {}_{l}g^{-1}=\varkappa^{-1ab}\mathcal{R}_{a}\otimes\mathcal{R}_{b}.
\label{eq_2.42}
\end{equation}

The Laplace-Beltrami operator corresponding to the Killing metric (\ref{eq_2.32}) is given by
\begin{equation}
\Delta=\gamma^{ab}\mathcal{L}_{a}\mathcal{L}_{b}=
\gamma^{ab}\mathcal{R}_{a}\mathcal{R}_{b}.\label{eq_2.43}
\end{equation}
Quite similarly, for the right-invariant metric (\ref{eq_2.39}) and
left-invariant metric (\ref{eq_2.40}) we would have respectively
\begin{equation}
{}_{r}\Delta=\varkappa^{ab}\mathcal{L}_{a}\mathcal{L}_{b},\qquad
{}_{l}\Delta=\varkappa^{ab}\mathcal{R}_{a}\mathcal{R}_{b}.\label{eq_2.44}
\end{equation}
Obviously, if $G$ is non-Abelian, these expressions are different
when $\varkappa_{ab}\neq\gamma_{ab}$.

One can show that all these expressions coincide with the usual definition
of the Laplace-Beltrami operator \cite{Godl}:
\begin{equation}
\Delta=g^{ab}\nabla_{a}\nabla_{b},\label{eq_2.45}
\end{equation}
where $\nabla_{a}$ denotes the Levi-Civita affine connection induced
by the corresponding vector tensors (\ref{eq_2.37}), (\ref{eq_2.39}),
(\ref{eq_2.40}). Obviously, this coincides with the analytical formula:
\begin{equation}
\Delta\psi=\frac{1}{\sqrt{\left|g\right|}}\underset{i,j}{\sum}\frac{\partial}{\partial k^{i}}\left(\sqrt{\left|g\right|}g^{ij}\frac{\partial\psi}{\partial k^{j}}\right),\label{eq_2.46}
\end{equation}
where again for $g$ the expressions (\ref{eq_2.37}), (\ref{eq_2.39}),
(\ref{eq_2.40}) are substituted, their contravariant inverses $g^{ij}$
are used, and $\left|g\right|$ denotes the determinant of the matrix
$\left[g_{ij}\right]$. 

The Haar measure in $G$ is identical with the $n$-form 
\begin{equation}
\mathcal{L}^{1}\wedge\ldots\wedge\mathcal{L}^{n}=
\mathcal{R}^{1}\wedge\ldots\wedge\mathcal{R}^{n}\label{eq_2.47}
\end{equation}
in the sense that
\begin{equation}
\int f(g)dg=\int f\mathcal{L}^{1}\wedge\ldots\wedge\mathcal{L}^{n}=\int f\mathcal{R}^{1}\wedge\ldots\wedge\mathcal{R}^{n}.\label{eq_2.48}
\end{equation}
In this prescription it is implicitly assumed that the orientation
of $G$ is chosen in such a way that the integral of non-negative
functions is non-negative. Analytically we have that
\begin{equation}
\int f(g)dg=\int f\det[\mathcal{L}^{a}{}_{i}]dk^{1}\ldots dk^{n}=
\int f\det[\mathcal{R}^{a}{}_{i}]dk^{1}\ldots dk^{n}.\label{eq_2.49}
\end{equation}
This integration coincides (up to a constant factor) with the usual
Riemann integration 
\begin{equation}
\int f(h)dh=\int f\sqrt{\left|g\right|}dk^{1}\ldots dk^{n},\label{eq_2.50}
\end{equation}
where $g$ denotes any of the metric tensors (\ref{eq_2.37}), (\ref{eq_2.39}),
(\ref{eq_2.40}). The Laplace-Beltrami operators (\ref{eq_2.43}),
(\ref{eq_2.44}), (\ref{eq_2.45}) are formally self-adjoint (symmetric) with respect to the usual scalar product in $L^{2}(G)$.

The properties (120)--(122) in \cite{1} imply immediately that 
\begin{equation}
\mathcal{L}_{a}\left(F*G\right)=\left(\mathcal{L}_{a}F\right)*G,\qquad
\mathcal{R}_{a}\left(F*G\right)=F*\left(\mathcal{R}_{a}G\right);\label{eq_2.51}
\end{equation}
again we conclude that $\mathcal{L}_{a}$, $\mathcal{R}_{a}$ are
not differentiations of the convolution algebra, although they are
so for the pointwise product algebra. If $F$ is constant on equivalence
classes of adjoint elements, i.e., if it is a linear combination or
series of idempotents $\varepsilon(\alpha)$ or characters
\begin{equation}\label{eq_2.51a}
\chi(\alpha)=\frac{1}{n(\alpha)}\varepsilon(\alpha),
\end{equation}
then, obviously, 
\begin{equation}
\mathcal{A}_{a}F=0,\label{eq_2.52}
\end{equation}
therefore, 
\begin{equation}
\mathcal{L}_{a}F=\mathcal{R}_{a}F.\label{eq_2.53}
\end{equation}
In particular, it is so for the Dirac distribution $\delta$ which
formally plays the role of the convolution unity. Let us stress that
in differential manifolds the distributions are well defined. In any case,
for any finite subset $I\subset\Omega$, 
\begin{equation}
\delta\left(I\right)=\underset{\alpha\in I}{\sum}\varepsilon(\alpha)\label{eq_2.54}
\end{equation}
is the well-defined unity of the two-sided ideal 
\begin{equation}
M(I):=\underset{\alpha\in I}{\otimes}M(\alpha).\label{eq_2.55}
\end{equation}
If $J$ is a family of finite subsets of $\Omega$ ordered by inclusion
and such that 
\begin{equation}
\underset{I\in J}{\bigcup}M(I)=\Omega,\label{eq_2.56}
\end{equation}
then $\delta$ is the distribution limit of the generalized sequence
$J\ni I\rightarrow\delta(I)$. 

Equations (\ref{eq_2.51}) imply that 
\begin{equation}
\mathcal{L}_{a}F=\mathcal{L}_{a}\left(\delta*F\right)=
\left(\mathcal{L}_{a}\delta\right)*F,\qquad \mathcal{R}_{a}F=
\mathcal{R}_{a}\left(F*\delta\right)=F*\left(\mathcal{R}_{a}\delta\right),
\label{eq_2.57}
\end{equation}
for any differentiable function $F$. This reduces separately to the
ideals $M(\alpha)$, where the action of operators $\mathcal{L}_{a}$, $\mathcal{R}_{a}$ reduces respectively to the left and right convolutions with $\mathcal{L}_{a}\varepsilon(\alpha)$, $\mathcal{R}_{a}\varepsilon(\alpha)$.

Let us quote some important and intuitive commutation relations in the
convolution algebra, 
\begin{eqnarray}
\left(\mathcal{L}_{a}\delta\right)*\left(\mathcal{L}_{b}\delta\right)-\left(\mathcal{L}_{b}\delta\right)*\left(\mathcal{L}_{a}\delta\right) & =&-C^{k}{}_{ab}\left(\mathcal{L}_{k}\delta\right),\label{eq_2.58a}\\
\left(\mathcal{R}_{a}\delta\right)*\left(\mathcal{R}_{b}\delta\right)-\left(\mathcal{R}_{b}\delta\right)*\left(\mathcal{R}_{a}\delta\right) & =&-C^{k}{}_{ab}\left(\mathcal{R}_{k}\delta\right).\label{eq_2.58b}
\end{eqnarray}
This is of course the same relation written in two ways, because $\mathcal{L}_{a}\delta=\mathcal{R}_{a}\delta$.

Roughly speaking, the functions constant on manifolds of mutually
adjoint elements are scalars of the group of inner automorphisms of
$G$, they satisfy the conditions 
\begin{equation}
\mathcal{A}_{c}F=0,\quad\textrm{e.g.},\quad \mathcal{A}_{c}\delta=0,\qquad
\mathcal{A}_{c}\underset{\alpha}{\sum}c_{\alpha}\varepsilon(\alpha)=0.
\label{eq_2.59}
\end{equation}
It is no longer the case with their $\mathcal{L}_{r}$-derivatives,
\begin{equation}
\mathcal{L}_{a}F=\mathcal{R}_{a}F,\quad \textrm{e.g.},\quad \mathcal{L}_{a}\delta=\mathcal{R}_{a}\delta,\qquad
\mathcal{L}_{a}\underset{\alpha}{\sum}c_{\alpha}\varepsilon(\alpha)=
\mathcal{R}_{a}\underset{\alpha}{\sum}c_{\alpha}\varepsilon(\alpha).
\label{eq_2.60}
\end{equation}
Roughly speaking, they are vectors of the group of inner automorphisms,
e.g., denoting
\begin{equation}
Q_{a}:=\mathcal{L}_{a}\delta=\mathcal{R}_{a}\delta,\label{eq_2.61}
\end{equation}
we have 
\begin{equation}
\mathcal{A}_{a}Q_{b}=-C^{k}{}_{ab}Q_{k},\label{eq_2.62}
\end{equation}
and similarly, for all other quantities in (\ref{eq_2.60}) and their
multiples by functions constant on equivalence classes. Similarly,
we have higher-order tensors, e.g., 
\begin{equation}
Q_{ab}=\mathcal{L}_{a}\mathcal{L}_{b}\delta=
\left(\mathcal{L}_{a}\delta\right)*\left(\mathcal{L}_{b}\delta\right)=
Q_{a}*Q_{b};\label{eq_2.63}
\end{equation}
they satisfy
\begin{equation}
\mathcal{A}_{c}Q_{ab}=-C^{k}{}_{ca}Q_{kb}-C^{k}{}_{cb}Q_{ak},\label{eq_2.64}
\end{equation}
and so on, for example, for 
\begin{equation}
Q_{abc}=\mathcal{L}_{a}\mathcal{L}_{b}\mathcal{L}_{c}\delta=Q_{a}*Q_{b}*Q_{c}
\label{eq_2.65}
\end{equation}
we have 
\begin{equation}
\mathcal{A}_{d}Q_{abc}=
-C^{k}{}_{da}Q_{kbc}-C^{k}{}_{db}Q_{akc}-C^{k}{}_{dc}Q_{abk},\label{eq_2.66}
\end{equation}
etc.

Casimir $\mathcal{L}$-operators are polynomials of $\mathcal{L}_{b}$
with constant coefficients, commuting with all $\mathcal{L}_{a}$.
They are expected to be polynomials of $\mathcal{L}_{b}$ with coefficients
built intrinsically of structure constants $C$, like (\ref{eq_2.4}),
(\ref{eq_2.34}), (\ref{eq_2.35}), (\ref{eq_2.36}) or rather their
versions with $\gamma$-raised indices. The most important example
is the Laplace-Beltrami operator (\ref{eq_2.43}); it is clear that
\begin{equation}
\left[\Delta,\mathcal{L}_{a}\right]=\left[\Delta,\mathcal{R}_{a}\right]=0.
\label{eq_2.67}
\end{equation}
Other expected quantities of this type are 
\begin{equation}
\gamma(m)^{i_{1}\ldots i_{m}}\mathcal{L}_{i_{1}}\ldots\mathcal{L}_{i_{m}},\label{eq_2.68}
\end{equation}
etc.; obviously, the raising of indices is meant in the sense of the
Killing tensor. In the group-algebraic representation, these Casimir
objects are given by functions/distributions like 
\begin{eqnarray}
C(2)&=&\gamma^{ij}\left(\mathcal{L}_{i}\delta\right)*\left(\mathcal{L}_{j}
\delta\right)=\gamma^{ij}\mathcal{L}_{i}\mathcal{L}_{j}\delta,\label{eq_2.69a}\\
C(m)&=&\gamma(m)^{i_{1}\ldots i_{m}}\left(\mathcal{L}_{i_{1}}\delta\right)*\cdots*\left(\mathcal{L}_{i_{m}}
\delta\right)=\gamma(m)^{i_{1}\ldots i_{m}}\mathcal{L}_{i_{1}}\ldots\mathcal{L}_{i_{m}}\delta.\qquad \label{eq_2.69b} \end{eqnarray}
They are expected to satisfy 
\begin{equation}
C(m)*f-f*C(m)=0\label{eq_2.70}
\end{equation}
(central elements of the convolution algebra).

To avoid distributions, one can consider their "$\alpha$-versions",
built of elements of $M(\alpha)$,
\begin{eqnarray}
C(2,\alpha)&= & \gamma^{ij}\left(\mathcal{L}_{i}\varepsilon(\alpha)\right)*
\left(\mathcal{L}_{j}\varepsilon(\alpha)\right),\label{eq_2.71a}\\ 
C(m,\alpha)&= & \gamma(m)^{i_{1}\ldots i_{m}}\left(\mathcal{L}_{i_{1}}\varepsilon(\alpha)\right)*\cdots*
\left(\mathcal{L}_{i_{m}}\varepsilon(\alpha)\right).\label{eq_2.71b}
\end{eqnarray}

Similarly, for any fixed $\alpha\in\Omega$, the quantities (\ref{eq_2.61}),
(\ref{eq_2.63}), (\ref{eq_2.65}), and so on become usual functions:
\begin{eqnarray}
Q_{a}(\alpha)&=& \mathcal{L}_{a}\varepsilon(\alpha)=\mathcal{R}_{a}\varepsilon(\alpha),
\label{eq_2.72a}\\
Q_{ab}(\alpha)&=& \mathcal{L}_{a}\mathcal{L}_{b}\varepsilon(\alpha)=\left(\mathcal{L}_{a}
\varepsilon(\alpha)\right)*\left(\mathcal{L}_{b}\varepsilon(\alpha)\right)=
Q_{a}(\alpha)*Q_{b}(\alpha),\label{eq_2.72b}\\
Q_{abc}(\alpha)&=&Q_{a}(\alpha)*Q_{b}(\alpha)*Q_{c}(\alpha),\label{eq_2.72c}\\
\vdots && \vdots \nonumber \\
Q_{ab\ldots r}(\alpha) &=& Q_{a}(\alpha)*Q_{b}(\alpha)*\cdots*Q_{r}(\alpha).\label{eq_2.72d}
\end{eqnarray}
etc. Obviously, $Q_{a}$, $Q_{ab}$, $Q_{abc}$, etc., are distributions
obtained as series (in the distribution sense of limit) of all the
above $Q$-s. One important circumstance must be stressed: The quantities
$Q_{ab\ldots r}$ are tensors under the action of automorphisms, however they
are not irreducible tensors, because they are not symmetric if $G$
is non-Abelian. To obtain irreducible tensors one must take their
symmetric parts, skew-symmetric ones, and remove the $\gamma$-traces
from the symmetric parts.

For any fixed $\alpha$ the tensors $Q_{a}$, $Q_{ab}$, etc.,
form some basis of $M(\alpha)$ alternative to $\varepsilon(\alpha)_{ij}$.
Obviously, when $\alpha$ is fixed, the order of tensors $Q(\alpha)$
terminates at some value, because $\dim M(\alpha)=n(\alpha)^{2}$
cannot be exceeded. 

From some point of view one might suppose that the pointwise products
of $Q_{a}$, e.g., $Q_{a}Q_{b}$, $Q_{a}Q_{b}Q_{c}$, etc., might
be simpler and more convenient. And they are tensors of $\mathcal{A}_{i}$
as well. However, it is not the case, because $Q_{a}(\alpha)Q_{b}(\alpha)$,
etc., are no longer elements of $M(\alpha)$. Nevertheless, they may
be useful in a sense. They may become elements of $M(\alpha)$ when
multiplied by appropriate scalars under inner automorphisms, i.e., multiplied by appropriate functions $f(\alpha)$ constant on classes
of adjoint elements, thus, satisfying (\ref{eq_2.52}). 

The matrices of irreducible representations $D(\alpha)$ will be represented
(at least locally, in some neighbourhood of the group identity), as
follows:
\begin{equation}
D(\alpha)(g)=\exp\left(k^{a}e(\alpha)_{a}\right),\qquad g\left(k^{1},\ldots,k^{n}\right)=\exp\left(k^{a}e_{a}\right), \label{eq_2.73}
\end{equation}
where, obviously, $e(\alpha)$ are $n(\alpha)\times n(\alpha)$ matrices
which obey the commutation rules (\ref{eq_2.1}):
\begin{equation}
\left[e(\alpha)_{a},e(\alpha)_{b}\right]=e(\alpha)_{k}C^{k}{}_{ab}.\label{eq_2.74}
\end{equation}
If $D(\alpha)$ are unitary, that is always assumed here, then $e(\alpha)$
are anti-Hermitian, so we have that 
\begin{equation}
D(\alpha)^{+}=D(\alpha)^{-1},\qquad e(\alpha)^{+}=-e(\alpha). \label{eq_2.75}
\end{equation}

In quantum-mechanical considerations the fundamental role is played
by Hermitian matrices
\begin{equation}
\Sigma(\alpha)_{a}=\frac{\hbar}{i}\: e(\alpha)_{a}=\Sigma(\alpha)_{a}^{+} \label{eq_2.76}
\end{equation}
which obey the commutation rules analogous to (\ref{eq_2.21a})--(\ref{eq_2.21c}):
\begin{equation}
\frac{1}{\hbar i}\left[\Sigma(\alpha)_{a},\Sigma(\alpha)_{b}\right]=
C^{k}{}_{ab}\Sigma(\alpha)_{k}.\label{eq_2.77}
\end{equation}
Then, obviously, we have the favourite formulas of physicists:
\begin{eqnarray}
D(\alpha)\left(g\left(\overline{k}\right)\right)&=&\exp \left(\frac{i}{\hbar}k^{a}\Sigma(\alpha)_{a}\right),\label{eq_2.78a}\\
L\left(g(\overline{k})^{-1}\right)&=& \exp \left(\frac{i}{\hbar}k^{a}\mathbf{\Sigma}_{a}\right),\label{eq_2.78b}\\
R\left(g(\overline{k})^{-1}\right)&=& \exp \left(\frac{i}{\hbar}k^{a}\mathbf{\widehat{\Sigma}}_{a}\right),\label{eq_2.78c}
\end{eqnarray}
obviously, the last two formulas are meant in an appropriate function domain, if to be meaningful. The representation property and definition of operators $\mathbf{\mathcal{L}}_{a}$, $\mathbf{\mathcal{R}}_{a}$, $\mathbf{\mathcal{A}}_{a}$ and their Hermitian counterparts $\mathbf{\Sigma}_{a}$, $\mathbf{\widehat{\Sigma}}_{a}$, $\mathbf{\Delta}_{a}$ imply that the matrix-valued functions $D(\alpha)$ on $G$ (equivalently $\varepsilon(\alpha)=n(\alpha)D(\alpha)$) satisfy the following differential equations:
\begin{eqnarray}
\mathbf{\mathcal{L}}_{a}D(\alpha)&=& e(\alpha)_{a}D(\alpha),\label{eq_2.79a}\\
\mathbf{\mathcal{R}}_{a}D(\alpha)&=& D(\alpha)e(\alpha)_{a},\label{eq_2.79b}\\
\mathbf{\mathcal{A}}_{a}D(\alpha)&=& e(\alpha)_{a}D(\alpha)-D(\alpha)e(\alpha)_{a}=
\left[e(\alpha)_{a},D(\alpha)\right],\label{eq_2.79c}
\end{eqnarray}
or, in terms of "Hermitian" operators,
\begin{eqnarray}
\mathbf{\Sigma}_{a}D(\alpha)&=& \Sigma(\alpha)_{a}D(\alpha),\label{eq_2.80a}\\
\mathbf{\widehat{\Sigma}}_{a}D(\alpha)&=& D(\alpha)\Sigma(\alpha)_{a},\label{eq_2.80b}\\
\mathbf{\Delta}_{a}D(\alpha)&=&\left[\Sigma(\alpha)_{a},D(\alpha)\right].
\label{eq_2.80c}
\end{eqnarray}

Let $C(\mathcal{L})$, $C(\mathcal{R})$, $C(\mathcal{A})$ denote the mentioned Casimir operators; let us remind that $C(\mathcal{L})$ commute with all $\mathcal{L}_{a}$-operators, $C(\mathcal{R})$ commute with all $\mathcal{R}_{a}$-operators, and $C(\mathcal{A})$ commute with all $\mathcal{A}_{a}$-operators. They are built in a polynomial way respectively of $\mathcal{L}$, $\mathcal{R}$, $\mathcal{A}$. Obviously, $C(\mathcal{L})$-Casimirs commute also with all $\mathcal{R}$- and $\mathcal{A}$-operators and $C(\mathcal{R})$-Casimirs also commute with $\mathcal{L}$- and $\mathcal{A}$-operators. This follows from the obvious fact that all $\mathcal{L}$-operators commute with all $\mathcal{R}$-operators. But attention: $C(\mathcal{A})$-Casimirs do not commute with all $\mathcal{L}$- and $\mathcal{R}$-operators. However, they do commute with $C(\mathcal{L})$- and $C(\mathcal{R})$-Casimirs. For physical reasons one uses often the $C(\mathbf{\Sigma})$-, $C(\mathbf{\widehat{\Sigma}})$-, and $C(\mathbf{\Delta})$-Casimirs. They are built of $\mathbf{\Sigma}$-, $\mathbf{\widehat{\Sigma}}$-, $\mathbf{\Delta}$-operators just like $C(\mathbf{\mathcal{L}})$, $C(\mathbf{\mathcal{R}})$, $C(\mathbf{\mathcal{A}})$ are built of the indicated operators. Usually there are a few ones of each kind; if necessary, some additional label is introduced (e.g., polynomial degree, etc.).

The use of differential operators acting on the functions on $G$, in particular, the use of their associative products, enables one to avoid dealing with more abstract and non-intuitive notion of the enveloping algebra of $G^{'}$.

The most important Casimirs are $\gamma$-quadratic functions of $\mathcal{L}$, $\mathcal{R}$, $\mathcal{A}$:
\begin{eqnarray}
C(\mathcal{L},2)=C(\mathcal{R},2)&=& \Delta = \gamma^{ab}\mathcal{L}_{a}\mathcal{L}_{b} = \gamma^{ab}\mathcal{R}_{a}\mathcal{R}_{b},\label{eq_2.81b}\\
C(\mathcal{A},2) &=& \gamma^{ab}\mathcal{A}_{a}\mathcal{A}_{b}.\label{eq_2.81c}
\end{eqnarray}
As mentioned, in addition to the obvious rules,
\begin{equation}
\left[C(\mathcal{L},2),\mathcal{L}_{a}\right]=
\left[C(\mathcal{L},2),\mathcal{R}_{a}\right]= \left[C(\mathcal{A},2),\mathcal{A}_{a}\right]=0,\label{eq_2.82}
\end{equation}
we have also
\begin{equation}
\left[C(\mathcal{L},2),\mathcal{A}_{a}\right]= \left[C(\mathcal{L},2),C(\mathcal{A},2)\right]=0. \label{eq_2.83}
\end{equation}

The corresponding expressions for "Hermitian" operators will be denoted by
\begin{equation}\label{eq_2.83a}
C\left(\mathbf{\Sigma},2\right)=C\left(\mathbf{\widehat{\Sigma}},2\right), \qquad C\left(\mathbf{\Delta},2\right), \qquad {\rm etc.}
\end{equation}
They are built according to the prescriptions for $C(\mathbf{\mathcal{L}},2)$, $C(\mathbf{\mathcal{R}},2)$, $C(\mathbf{\mathcal{A}},2)$ with $\mathbf{\Sigma}$, $\mathbf{\widehat{\Sigma}}$, $\mathbf{\Delta}$ substituted respectively instead of $\mathbf{\mathcal{L}}$, $\mathbf{\mathcal{R}}$, $\mathbf{\mathcal{A}}$, therefore, for quadratic Casimirs we have
\begin{equation}
C\left(\mathbf{\Sigma},2\right)= -\hbar^{2}C (\mathcal{L},2),\quad C\left(\mathbf{\widehat{\Sigma}},2\right)=-\hbar^{2}C (\mathbf{\mathcal{R}},2), 
\quad C\left(\mathbf{\Delta},2\right)=-\hbar^{2}C (\mathbf{\mathcal{A}},2) \label{eq_2.81a}
\end{equation}
and similarly for other Casimirs.

When we fix some $\alpha$ and act with our Casimirs on functions $D(\alpha)$ $\left(\varepsilon(\alpha)\right)$, they simply suffer the multiplication by scalars, just the eigenvalues of Casimirs. This follows from the Schur lemma, because $D(\alpha)$ are irreducible. Therefore, e.g., iterating appropriately (\ref{eq_2.79a})--(\ref{eq_2.79c}), (\ref{eq_2.80a})--(\ref{eq_2.80c}), we obtain
\begin{eqnarray}
\gamma^{ab}\mathbf{\mathcal{L}}_{a}\mathbf{\mathcal{L}}_{b}D(\alpha)=
\gamma^{ab}\mathcal{R}_{a}\mathcal{R}_{b}D(\alpha)&=& \gamma^{ab}e(\alpha)_{a}e(\alpha)_{b}D(\alpha)= C(2,\alpha)D(\alpha),\qquad\ \label{eq_2.84a} \\
\gamma^{ab}\mathbf{\Sigma}_{a}\mathbf{\Sigma}_{b}D(\alpha)=
\gamma^{ab}\mathbf{\widehat{\Sigma}}_{a}\mathbf{\widehat{\Sigma}}_{b}D(\alpha) &=&-\hbar^{2} C(2,\alpha)D(\alpha),\label{eq_2.84b}
\end{eqnarray}
where
\begin{equation}
\gamma^{ab}e(\alpha)_{a}e(\alpha)_{b}=C(2,\alpha){\rm Id}_{n(\alpha)} \label{eq_2.85}
\end{equation}
and $C(2,\alpha)$ are elements of the spectrum of $\Delta$ (\ref{eq_2.43}). These eigenvalues are $n(\alpha)^{2}$-fold degenerate. It was mentioned that although $\gamma^{ab}\mathcal{A}_{a}\mathcal{A}_{b}$ does not commute in general with $\mathcal{L}_{a}$, $\mathcal{R}_{b}$, nevertheless, it does commute with $\Delta= \gamma^{ab}\mathcal{L}_{a}
\mathcal{L}_{b}= \gamma^{ab}\mathcal{R}_{a}\mathcal{R}_{b}$. However, $D(\alpha)_{ij}$ are not their common eigenfunctions; indeed
\begin{equation}
\gamma^{ab}\mathcal{A}_{a}\mathcal{A}_{b}D(\alpha)=2C(2,\alpha)D(\alpha)- 2\gamma^{ab} e(\alpha)_{a} D(\alpha) e(\alpha)_{b}, \label{eq_2.86}
\end{equation}
i.e.,
\begin{equation}
\gamma^{ab}\Delta_{a}\Delta_{b}D(\alpha)=-2C(2,\alpha)\hbar^{2} D(\alpha)- 2\gamma^{ab} \Sigma(\alpha)_{a} D(\alpha) \Sigma(\alpha)_{b}. \label{eq_2.87}
\end{equation}
Nevertheless, their common eigenfunctions do exist and are given by (\ref{eq_2.72a})--(\ref{eq_2.72d}); for any fixed $\alpha \in \Omega$, the order of tensors (\ref{eq_2.72a})--(\ref{eq_2.72d}) terminates at some fixed value.

It is seen that in the action on functions $\varepsilon(\alpha)_{ij} = n(\alpha)D(\alpha)_{ij}$, our differential operators become algebraic. This is just the obvious counterpart and generalization of the well-known facts in Fourier analysis. Let us quote a few obvious and practically important formulas. 

It was mentioned earlier about the Peter-Weyl expansion (102) in \cite{1}; let us write it a bit more symbolically as
\begin{equation}
F=\sum_{\alpha \in \Omega}{\rm Tr}\left(F(\alpha)^{T} \varepsilon(\alpha)\right)=\sum_{\alpha \in \Omega}{\rm Tr}\left(F(\alpha)^{T} D(\alpha)\right)n(\alpha). \label{eq_2.88}
\end{equation}

The general operations of group algebras are then represented in a suggestive way by the corresponding operations performed on the matrices $F(\alpha)$, cf. (104)--(107) in \cite{1}. Together with the formulas (\ref{eq_2.79a})--(\ref{eq_2.79c}), (\ref{eq_2.80a})--(\ref{eq_2.80c}), (\ref{eq_2.84a})--(\ref{eq_2.84b}), (\ref{eq_2.86}), (\ref{eq_2.87}) this implies that the action of differential operators may be expressed in the following way by the corresponding algebraic operations on the representing matrices $F(\alpha)$:
\begin{eqnarray}
\mathbf{\mathcal{L}}_{a}, \mathbf{\Sigma}_{a}: &\quad & F(\alpha)\mapsto e(\alpha)_{a}^{T}F(\alpha), \qquad F(\alpha)\mapsto \Sigma(\alpha)_{a}^{T}F(\alpha),\label{eq_2.89a} \\
\mathbf{\mathcal{R}}_{a}, \mathbf{\widehat{\Sigma}}_{a}: &\quad & F(\alpha)\mapsto F(\alpha)e(\alpha)_{a}^{T}, \qquad F(\alpha)\mapsto F(\alpha) \Sigma(\alpha)_{a}^{T},\label{eq_2.89b} \\
\mathbf{\mathcal{A}}_{a}, \mathbf{\Delta}_{a}: &\quad & F(\alpha)\mapsto \left[e(\alpha)_{a}^{T},F(\alpha)\right], \qquad F(\alpha)\mapsto \left[\Sigma(\alpha)_{a}^{T},F(\alpha)\right].\qquad \label{eq_2.89c}
\end{eqnarray}
Therefore, the action of $\gamma^{ab}\mathcal{L}_{a}\mathcal{L}_{b}=
\gamma^{ab}\mathcal{R}_{a}\mathcal{R}_{b}= \Delta$ is represented by multiplication of matrices $F(\alpha)$ by $C(2, \alpha)$; and similarly for other Casimirs. 

Let us mention that for some purposes the convention of transposed $F(\alpha)$-matrices might be more convenient, namely,
\begin{equation}
F= \sum_{\alpha \in \Omega}{\rm Tr}\left(F(\alpha)\varepsilon(\alpha)\right)= \sum_{\alpha \in \Omega}{\rm Tr}\left(F(\alpha)D(\alpha)\right)n(\alpha).\label{eq_2.90}
\end{equation}
A disadvantage is that then $F \ast G$ is not represented by the system of $F(\alpha)G(\alpha)$ but $G(\alpha)F(\alpha)$. But, and this is an aesthetic advantage, the matrix transposition is avoided, namely, the $\mathbf{\mathcal{L}}_{a}/\mathbf{\Sigma}_{a}$ act respectively as follows:
\begin{eqnarray}
\mathbf{\mathcal{L}}_{a},\mathbf{\Sigma}_{a}: &\quad & F(\alpha)\mapsto F(\alpha)e(\alpha)_{a}, \quad F(\alpha)\mapsto F(\alpha) \Sigma(\alpha)_{a}, 
\label{eq_2.91a}\\
\mathbf{\mathcal{R}}_{a},\mathbf{\widehat{\Sigma}}_{a}: &\quad & F(\alpha)\mapsto R(\alpha)_{a}F(\alpha), \quad F(\alpha)\mapsto  \Sigma(\alpha)_{a}F(\alpha), 
\label{eq_2.91b}\\
\mathbf{\mathcal{A}}_{a},\mathbf{\Delta}_{a}: &\quad & F(\alpha)\mapsto \left[F(\alpha),e(\alpha)_{a}\right], \quad F(\alpha)\mapsto \left[F(\alpha),\Sigma(\alpha)_{a}\right].\label{eq_2.91c}
\end{eqnarray}
But again, a disadvantage is that the left/right differential generators are represented algebraically by the right/left matrix multiplication, thus, conversely. Of course, all this is a merely matter of convention.

If we use the convention (\ref{eq_2.88}), then the functions (\ref{eq_2.61}), (\ref{eq_2.63}), (\ref{eq_2.65}), etc., i.e.,
\begin{equation}
{}^{l}Q_{ab \ldots k}= \mathcal{L}_{a}\mathcal{L}_{b} \ldots \mathcal{L}_{k} \delta = \left(\mathcal{L}_{a} \delta \right) \ast \cdots \ast \left(\mathcal{L}_{k} \delta \right) = Q_{a} \ast Q_{b} \ast \cdots \ast Q_{k} 
\label{eq_2.92}
\end{equation}
are represented by matrices
\begin{equation}
{}^{l}\widehat{Q}(\alpha)_{ab \ldots k} = e(\alpha)_{a}{}^{T} e(\alpha)_{b}{}^{T} \ldots e(\alpha)_{k}{}^{T}. \label{eq_2.93}
\end{equation}
And similarly, the functions
\begin{eqnarray}
{}^{r}Q_{ab \ldots k}&=& 
\mathcal{R}_{a}\mathcal{R}_{b} \ldots \mathcal{R}_{k} \delta = \mathcal{L}_{k} \ldots \mathcal{L}_{b}\mathcal{L}_{a} \delta = \left(\mathcal{R}_{a} \delta \right) \ast \left(\mathcal{R}_{b} \delta \right) \ast \ldots \ast \left(\mathcal{R}_{k} \delta \right)\nonumber\\
&=& \left(\mathcal{L}_{k} \delta \right) \ast \ldots \ast \left(\mathcal{L}_{b} \delta \right) \ast \left(\mathcal{L}_{a} \delta \right)
=Q_{k} \ast \ldots \ast Q_{b} \ast Q_{a} = {}^{l}Q_{k \ldots ba}\qquad\quad \label{eq_2.94}	
\end{eqnarray}
are represented by matrices
\begin{equation}
{}^{r}\widehat{Q}(\alpha) = e(\alpha)_{k}{}^{T}  \ldots e(\alpha)_{b}{}^{T} e(\alpha)_{a}{}^{T}. \label{eq_2.95}
\end{equation}
If we use the convention (\ref{eq_2.90}), then instead of (\ref{eq_2.93}), (\ref{eq_2.95}) we obtain respectively
\begin{eqnarray}
{}^{l}\widehat{Q}(\alpha)_{ab \ldots k} &=& e(\alpha)_{k} \ldots e(\alpha)_{b}  e(\alpha)_{a},\label{eq_2.96}\\
{}^{r}\widehat{Q}(\alpha)_{ab \ldots k} &=& e(\alpha)_{a} e(\alpha)_{b} \ldots e(\alpha)_{k}.\label{eq_2.97}
\end{eqnarray}

The Hermitian version of $Q_{a}$, representing a physical observable, is obtained by replacing the operators $\mathcal{L}_{a}$, $\mathcal{R}_{a}$ by (\ref{eq_2.20}), i.e., by 
\begin{equation}\label{eq_2.97a}
\mathbf{\Sigma}_{a}=\frac{\hbar}{i}\mathcal{L}_{a},\qquad \widehat{\mathbf{\Sigma}}_{a}=\frac{\hbar}{i}\mathcal{R}_{a}. 
\end{equation}
They are given by
\begin{equation}
\Sigma_{a}=\frac{\hbar}{i}Q_{a}; 
\label{eq_2.98}
\end{equation}
${Q}_{a}$ themselves are anti-Hermitian. 

Let us observe that if $G$ is non-Abelian (and here we concentrate mainly on semi-simple ones), then in general the functions ${}^{l}Q_{ab \ldots k}$, ${}^{r}Q_{ab \ldots k}$ and the representing matrices ${}^{l}\widehat{Q}(\alpha)_{ab \ldots k}$, ${}^{r}\widehat{Q}(\alpha)_{ab \ldots k}$ fail to be anti-Hermitian. Therefore, the corresponding monomials of $\Sigma(\alpha)$, $\Sigma(\alpha)^{T}$ are not Hermitian. But obviously, their symmetrizations
\begin{equation}
\Sigma(\alpha)_{(a} \ldots \Sigma(\alpha)_{k)}, \quad \Sigma(\alpha)^{T}_{(a} \ldots \Sigma(\alpha)^{T}_{k)} \label{eq_2.99}
\end{equation}
are Hermitian and so are the functions
\begin{equation}
\Sigma_{(a \ldots k)} = \left(\frac{\hbar}{i}\right)^{p}\left({\mathcal{L}}_{(a} \delta \right) \ast \cdots \ast \left({\mathcal{L}}_{k)} \delta \right)= \left(\frac{\hbar}{i}\right)^{p}\left({\mathcal{R}}_{(a} \delta \right) \ast \cdots \ast \left({\mathcal{R}}_{k)} \delta \right), \label{eq_2.100}
\end{equation}
where $p$ is the order of tensors (the number of convolution factors). Obviously, (\ref{eq_2.99}) are matrices of (\ref{eq_2.100}) when the conventions (\ref{eq_2.90}), (\ref{eq_2.88}) are used, respectively.

In realistic dynamical models Hamiltonians are usually given by simple algebraic functions of the above Hermitian elements of group algebras. As a rule, those Hamiltonians or their important terms are low-order polynomials. In special cases of high symmetry they are built according to the Casimir prescriptions.

Let us finish with some remarks concerning Abelian Lie groups. Obviously, the only (up to isomorphism) connected Abelian groups are: $\mathbb{R}^{n}$, $T^{n}=U(1)^{n}=\mathbb{R}^{n}/\mathbb{Z}^{n}$, and their Cartesian products $\mathbb{R}^{n} \times T^{m} $, i.e., linear spaces, tori and cylinders. The group operation in $\mathbb{R}^{n}$ is obviously meant as the addition of vectors (null vector being the neutral element); in $T^{n}$ it is meant as the quotient action obtained when $\mathbb{R}^{n}$ is divided by the "crystallographic" lattice $\mathbb{Z}^{n} \subset \mathbb{R}^{n}$.

Some conflicts between the above notational conventions and various customs from the classical Fourier analysis appear, so one must be careful with an automatic use of traditional formulas.

It is perhaps convenient to write down some formulas concerning $\mathbb{R}^{n}$ in the language of abstract vector space. So, let $V$ be a finite-dimensional linear space, and $V^{\ast}$ be its dual; we put $n= \dim V = \dim V^{\ast}$. We consider them as Abelian additive Lie groups. So, $G=V$ with the "$+$" composition rule, $\widehat{G}$ is isomorphic with $V^{\ast}$. And the particular choice of this isomorphism is a matter of convention. $G$ being non-compact, there is no standard of normalization. If $V$ is endowed with some fixed metric tensor $\gamma \in V^{\ast} \otimes V^{\ast}$, as it usually is in physical applications, then, of course, the standard of Lebesgue measure is fixed,
\begin{equation}
\int f(x) d\mu (x) = \int f e^{1}\wedge \ldots \wedge e^{n}, \label{eq_2.101}
\end{equation}
where $\left(\ldots, e^{a}, \ldots\right)$ is an arbitrary orthonormal co-basis in $V^{\ast}$:
\begin{equation}
g = \delta_{ij} e^{i} \otimes e^{j}. \label{eq_2.102}
\end{equation}
In arbitrary coordinates, including  curvilinear ones, we have
\begin{equation}
\int f(x) d\mu (x) = \int f(x)\sqrt{\det \left[g_{ij}\right]} \ dx^{1} \ldots dx^{n}. \label{eq_2.103}
\end{equation}
The dual linear space $V^{\ast}$ parametrizes the dual group $\widehat{V}$ with the help of the standard covering homomorphism of the (additive) $\mathbb{R}$ onto (multiplicative) $U(1)$,
\begin{equation}
\mathbb{R} \ni \varphi \mapsto \exp (i \varphi) \in U(1), \label{eq_2.104}
\end{equation}
so, $\chi\left(\underline{k}\right) \in \widehat{V}$ is given by
\begin{equation}
\left\langle \chi\left(\underline{k}\right), \overline{x} \right\rangle= \exp \left(i \left\langle \underline{k} , \overline{x} \right\rangle \right), \label{eq_2.105}
\end{equation}
where, obviously, $\left\langle \underline{k} , \overline{x} \right\rangle$ is the evaluation of $\underline{k} \in V^{\ast}$ on $\overline{x} \in V$; analytically
\begin{equation}
\left\langle \underline{k} , \overline{x} \right\rangle = k_{a}x^{a}. \label{eq_2.106}
\end{equation}
Using the language of quantum momentum $\underline{p}= \hbar \underline{k}$, one writes also
\begin{equation}
\left\langle \chi\left[\underline{p}\right], \overline{x} \right\rangle= \exp \left(\frac{i}{\hbar} \left\langle \underline{p} , \overline{x} \right\rangle \right) = \exp \left(\frac{i}{\hbar} \ p_{a}x^{a}\right). \label{eq_2.107}
\end{equation}
The corresponding conventions of Fourier analysis, particularly popular in quantum mechanics, are as follows:
\begin{eqnarray}
f\left(\overline{x}\right) &=& \frac{1}{(2\pi)^{n}} \int \widehat{f}\left(\underline{k}\right) \exp \left(i \left\langle \underline{k}, \overline{x} \right\rangle \right)d_{n}k \nonumber\\
& =&\frac{1}{(2\pi \hbar)^{n}} \int \widehat{f}\left(\underline{p}\right) \exp \left(\frac{i}{\hbar} \left\langle \underline{p}, \overline{x} \right\rangle \right)d_{n}p,\label{eq_2.108a} \\
\widehat{f}\left(\underline{k}\right) = \widehat{f}\left[\underline{p}\right] &=& \int f\left(\overline{x}\right) \exp \left(-\frac{i}{\hbar} \left\langle \underline{p}, \overline{x} \right\rangle \right)d_{n}x.\label{eq_2.108b}
\end{eqnarray}
The convolution on $V$ is meant in the usual convention,
\begin{equation}
\left(A \ast B\right)\left(\overline{x}\right)= \int A(\overline{y}) B(\overline{x}-\overline{y})d\overline{y}. \label{eq_2.109}
\end{equation}
We have then the following rules,
\begin{eqnarray}
\chi\left(\underline{k}\right) \ast \chi\left(\underline{l}\right)&=& (2\pi)^{n}\delta\left(\underline{k}-
\underline{l}\right)\chi\left(\underline{k}\right)= (2\pi)^{n}\delta\left(\underline{k}-
\underline{l}\right)\chi\left(\underline{l}\right),\label{eq_2.110a}\\
\left(\chi\left(\underline{k}\right), \chi\left(\underline{l}\right)\right)&=& (2\pi)^{n}\delta\left(\underline{k}-\underline{l}\right),\label{eq_2.110b}\\
\chi\left[\underline{p}\right] \ast \chi\left[\underline{p'}\right]&=& (2\pi \hbar)^{n}\delta\left(\underline{p}-
\underline{p'}\right)\chi\left[\underline{p}\right]= (2\pi \hbar)^{n}\delta\left(\underline{p}-
\underline{p'}\right)\chi\left[\underline{p'}\right],\qquad \label{eq_2.111a}\\
\left(\chi\left[\underline{p}\right], \chi\left[\underline{p'}\right]\right)&=& (2\pi \hbar)^{n}\delta\left(\underline{p}-\underline{p'}\right),
\label{eq_2.111b}
\end{eqnarray}
rather unpleasant ones, because of the $(2\pi)^{n}$-, $(2\pi \hbar)^{n}$-factors. But this has to do with the use of traditional symbols of analysis. If we remember that it is not $d_{n}k$, or $d_{n}p$, but rather $d_{n}k/(2\pi)^{n}$, $d_{n}p/(2\pi \hbar)^{n}$ that is a measure Fourier-synchronized with $d_{n}x$, that it is just $(2\pi )^{n}\delta\left(\underline{k}-\underline{l}\right)$ or $(2\pi \hbar)^{n}\delta\left(\underline{p}-\underline{p'}\right)$ that is to be interpreted as a "true Dirac delta", let us say $\Delta\left(\underline{k}-\underline{k'}\right)$, $\Delta\left(\underline{p}-\underline{p'}\right)$, respectively in the spaces of wave co-vectors and linear momenta.

There are various conventions concerning Fourier transforms and synchronization of measures on $G$, $\widehat{G}$, it is even stated in the book by Loomis \cite{2}, that it is "an interesting and non-trivial problem".

In classical analysis one often prefers the "symmetric" convention:
\begin{eqnarray}
A\left(\overline{x}\right) &=&  \frac{1}{(2\pi)^{n/2}} \int \widehat{A}\left(\underline{k}\right) \exp \left(i \left\langle \underline{k}, \overline{x} \right\rangle \right)d_{n}\underline{k},\label{eq_2.112a} \\
\widehat{A}\left(\underline{k}\right) &=& \frac{1}{(2\pi)^{n/2}} \int A\left(\overline{x}\right) \exp \left(-i \left\langle \underline{k}, \overline{x} \right\rangle \right)d_{n}\overline{x}.\label{eq_2.112b}
\end{eqnarray}
An additional advantage of this convention is that the iteration of Fourier transformation results in the inversion (total reflection) of the original function, with respect to the origin:
\begin{equation}
\widehat{\widehat{A}{\,}}(x)=A(-x). \label{eq_2.113}
\end{equation}
And, roughly speaking, Gauss function is invariant under Fourier transformation. More precisely, we have
\begin{equation}
\mathcal{G}(\overline{x}) = \exp \left(-\frac{1}{2}\ \overline{x} \cdot \overline{x} \right), \qquad \mathcal{\widehat{G}}(\underline{k}) = \exp\left(-\frac{1}{2} \ \underline{k} \cdot \underline{k} \right), \label{eq_2.114}
\end{equation}
where the scalar product in $V$ is meant in the sense of metric $g \in V^{\ast} \otimes V^{\ast}$, and in $V^{\ast}$ --- under its contravariant inverse $g^{-1} \in V \otimes V$,
\begin{equation}\label{eq_2.114a}
\overline{x} \cdot \overline{x} = g(x,x)=g_{ij}x^{i}x^{j}, \qquad \underline{k} \cdot \underline{k} = \widehat{g}(\underline{k},\underline{k})=g^{ij}x_{i}x_{j}. \end{equation}
If we identify $V=\mathbb{R}^{n}=V^{\ast}$, then the Gauss function is literally invariant under the Fourier transformation.

The counterparts of Clebsch-Gordon series (167), (174) in \cite{1} are very simple now, because
\begin{eqnarray}
\chi(\underline{k})\chi(\underline{l})&=&\chi(\underline{k}+\underline{l}),
\label{eq_2.115a}\\ \chi\left[\underline{p}\right]\chi\left[\underline{p'}\right]&=&
\chi\left[\underline{p}+\underline{p'}\right],\label{eq_2.115b} \\
\chi(\underline{k})\chi(\underline{l})&=& \int \delta(\underline{k}+\underline{l}-
\underline{m})\chi(\underline{m})d_{n}\underline{m},\label{eq_2.115c} \\
\chi\left[\underline{p}\right]\chi\left[\underline{p'}\right]&=& \int \delta(\underline{p}+\underline{p'}-
\underline{\pi})\chi\left[\underline{\pi}\right]d_{n}\underline{\pi}.
\label{eq_2.115d}
\end{eqnarray}

Let us now fix some symbols concerning the compact case $T^{n}=U(1)^{n}$. Just like $\mathbb{R}^{n}$ is an analytical model of any $n$-dimensional linear space over reals, $T^{n}$ is parametrized by the system of angles $\left(\varphi^{1}, \ldots, \varphi^{n} \right)$ taken modulo $2\pi$, or uniquely, by the system of unimodular complex numbers $\left(\zeta^{1}, \ldots, \zeta^{n} \right)$, $\zeta^{a}= \exp \left(i \varphi^{a}\right)$. Sometimes the convention "modulo $1$" is accepted instead "modulo $2\pi$", i.e., one puts $\zeta^{a}= \exp \left(i2\pi \xi^{a}\right)$. This is often used when $T^{n}$ is realized as a quotient of $V$ modulo the "crystallographic lattice" generated freely by some fixed basis $\left(\ldots, e_{a}, \ldots\right)$ in $V$. Obviously, that discrete translation group is isomorphic with $\mathbb{Z}^{n}$. The parametrization modulo $2\pi$ is more popular in theory of Fourier series. Torus is compact and it is natural to take the Haar measure normalized to unity, as usual. If the multiple Fourier series on $T^{n}$ are meant in the convention
\begin{equation}
f(\overline{\varphi})= \sum_{\underline{m}\in\mathbb{Z}^{n}} \widehat{f}(\underline{m}) \exp \left(i \underline{m} \cdot \overline{\varphi} \right), \label{eq_2.116}
\end{equation}
then the inverse formula for coefficients $\widehat{f}$ reads
\begin{equation}
\widehat{f}(\underline{m})= \frac{1}{(2\pi)^{n}} \int f(\overline{\varphi}) 
\exp \left(-i \underline{m} \cdot \overline{\varphi} \right)d_{n}\overline{\varphi}. \label{eq_2.117}
\end{equation}
Concerning notation, analytical meaning of the expressions above is as follows:
\begin{equation}
\underline{m}=\left(m_{1}, \ldots, m_{n}\right) \in \mathbb{Z}^{n}, \qquad \overline{\varphi}= \left(\varphi^{1}, \ldots, \varphi^{n}\right)^{T}, \label{eq_2.118}
\end{equation}
contractions in exponents are given by
\begin{equation}
\underline{m}\cdot \overline{\varphi}=m_{a}\varphi^{a}= m_{1}\varphi^{1} + \cdots + m_{n}\varphi^{n}, \label{eq_2.119}
\end{equation}
and the range of variables $\varphi^{a}$ in the integration element 
\begin{equation}\label{eq_2.119a}
d_{n}\overline{\varphi}= d\varphi^{1} \ldots d\varphi^{n}
\end{equation}
is given by $[0,2\pi]$.

It is seen that the occurrence of factors $(2\pi)^{-n}$ is reciprocal to that in Fourier analysis on $\mathbb{R}^{n}$. This spoils the formal analogy, but suits the convention that the Haar volume of compact groups equals the unity. To save the analogy, we would have to replace (\ref{eq_2.108a})--(\ref{eq_2.108b}) by 
\begin{eqnarray}
f(\overline{x}) &=& \int \widehat{f}(\underline{k}) \exp \left(i\left\langle  \underline{k}, \overline{x}\right\rangle \right) d_{n}\underline{k},\label{eq_2.120a} \\
\widehat{f}(\underline{k}) &=& \frac{1}{(2\pi)^{n}}\int f(\overline{x}) \exp \left(-i\left\langle  \underline{k}, \overline{x}\right\rangle \right) d_{n}\overline{x},\label{eq_2.120b}
\end{eqnarray}
which, by the way, is sometimes used indeed, however, it is incompatible with some other customs of physicists and their taste.

Characters on $T^{n}$ are labelled by multi-indices $\underline{m}\in\mathbb{Z}^{n}$,
\begin{equation}
\left\langle \chi(\underline{m}), \zeta(\overline{\varphi})\right\rangle =\left(\zeta^{1}\right)^{m_{1}} \ldots \left(\zeta^{n}\right)^{m_{n}}= \exp \left(i \underline{m} \cdot \overline{\varphi}\right). \label{eq_2.121}
\end{equation}
The idempotence and independence property is literally satisfied, because $T^{n}$ is compact and $\mathbb{Z}^{n}$ is discrete:
\begin{eqnarray}
\chi(\underline{m}) \ast \chi(\underline{l})&=& \delta_{\underline{m}\underline{l}}\chi(\underline{m}) = \delta_{\underline{m}\underline{l}}\chi(\underline{l})\label{eq_2.122}\\
\chi(\underline{m}) \chi(\underline{l})&=& \chi(\underline{m}+ \underline{l}) \label{eq_2.123}\\
\left(\chi(\underline{m}), \chi(\underline{l})\right)&=& \delta_{\underline{m}\underline{l}}, \label{eq_2.124}
\end{eqnarray}
where, obviously, the multi-index Kronecker symbol $\delta_{\underline{m}\underline{l}}$ vanishes if $\underline{m} \neq \underline{l}$ (i.e., at least one component of $\underline{m}$ differs from the corresponding component of $\underline{l}$), and $\delta_{\underline{m}\underline{l}}=1$ when $\underline{m} = \underline{l}$. In other words
\begin{equation}
\delta_{\underline{m}\underline{l}}= \delta_{m_{1}l_{1}} \ldots \delta_{m_{n}l_{n}}. \label{eq_2.125}
\end{equation}
Concerning the "Clebsch-Gordon" rule (\ref{eq_2.123}), its representation in terms of (167), (174) in \cite{1} reads
\begin{equation}
\chi(\underline{m}) \chi(\underline{l})= \sum_{\underline{\pi} \in \mathbb{Z}^{n}} \left(\underline{m} \ \underline{l} | \underline{\pi} \right) \left(\underline{m} \ \underline{l} | \underline{\pi} \right) \chi(\underline{\pi}), \label{eq_2.126}
\end{equation}
where
\begin{equation}
\left(\underline{m} \ \underline{l} | \underline{\pi} \right)= \delta_{\underline{m} + \underline{l} , \underline{\pi}}= \left(\underline{m} \ \underline{l} | \underline{\pi} \right)^{2}. \label{eq_2.127}
\end{equation}
Let us notice that in the non-compact case $G=\mathbb{R}^{n}$, the counterpart of (167) in \cite{1}, i.e., the right-hand side of (164) in \cite{1}, fails because the square of Dirac-delta is not well defined.

Obviously, if we take as an arena of our physics the discrete group $\mathbb{Z}^{n}$, then its dual group $T^{n}$ is compact and continuous. Again the mentioned problems with squared delta-distribution appear.

For certain reasons, first of all ones concerning quasiclassical analysis, it is interesting to discuss certain byproducts of the group structure in $G$. It is well known that the Lie algebra $G'$ of $G$ encodes a great amount of information about the global structure of $G$, although, of course, not the total information. This is due to the very analytic structure of Lie groups. Making use of exponential mapping of $G'$ into $G$ (not "onto" in general) one can "pull back" some structures of $G$ and some physics in $G$ to its tangent space $G'=T_{e}G$. But now, $G'$ as a finite-dimensional linear space is an Abelian Lie group under addition of its elements. Therefore, we can consider some physics, using the group algebra of $G'$ as an additive group of vectors. But of course this would be completely non-physical and non-interesting if we did not take into account the Lie-algebraic structure of $G^{\prime}$. This structure leads to certain additional structures and relationships in the group algebra of $G^{\prime}$. Namely, it is well know that the co-algebra $G^{\prime\ast}$, i.e., the algebraic dual space of $G^{\prime}$, carries the canonical Poisson structure. Namely, Poisson bracket of differentiable functions $A$, $B$ on $G^{\prime\ast}$ is analytically given by
\begin{equation}\label{eq_2.128}
\{A,B\}:=\sigma_{k}C^{k}{}_{lm}\frac{\partial A}{\partial \sigma_{l}}
\frac{\partial B}{\partial \sigma_{m}},
\end{equation}
where $\sigma_{k}$ are linear coordinates in $G^{\prime\ast}$ and $C^{k}{}_{lm}$ are structure constants with respect to these coordinates, or more precisely, with respect to the dual linear coordinates in $G^{\prime}$. Notice that, being linear functions on $G^{\prime\ast}$, i.e., elements of the second dual $G^{\prime\ast\ast}$, functions $\sigma_{k}$ are canonically identical with some basis vectors $e_{k}$ in $G^{\prime}$ and
\begin{equation}\label{eq_2.129}
\left[e_{l},e_{m}\right]=e_{k}C^{k}{}_{lm}.
\end{equation}
We might simply use the symbols $\sigma_{k}$ instead of $e_{k}$ in this formula, however, this might be perhaps a bit confusing, although essentially true.

It is obvious that the expression (\ref{eq_2.128}) is correct, i.e., coordinate-independent. It is well known that it may be formulated without any use of coordinates. Namely, take differentials $dA_{\sigma}$, $dB_{\sigma}$ at the point $\sigma\in G^{\prime\ast}$. Being linear functions on $G^{\prime\ast}\simeq T_{\sigma}G^{\prime\ast}$, they are canonically identical with some elements of $G^{\prime}$. We take their bracket/commutator $\left[dA_{\sigma},dB_{\sigma}\right]\in G^{\prime}$ and evaluate the one-form $\sigma\in G^{\prime\ast}$ on this vector, $\left\langle\sigma,\left[dA_{\sigma},dB_{\sigma}\right]\right\rangle$. One obtains the prescription assigning a number to any point $\sigma\in G^{\prime\ast}$. The resulting function is just the value of $\{A,B\}$ at $\sigma$:
\begin{equation}\label{eq_2.130}
\{A,B\}(\sigma)=
\left\langle\sigma,\left[dA_{\sigma},dB_{\sigma}\right]\right\rangle.
\end{equation}
The skew-symmetry is obvious and the Jacobi identity follows from the identity satisfied by structure constants, thus, finally from the Jacobi identity in Lie algebra.

Obviously, (\ref{eq_2.130}), (\ref{eq_2.128}) is defined only for differentiable functions. The associative algebra of smooth functions $C^{\infty}\left(G^{\prime\ast}\right)$ in the sense of pointwise product becomes simultaneously an infinite-dimensional Lie algebra under Poisson bracket. The two structures are compatible in the sense that the Poisson-bracket ${\rm ad}$-operation is a differentiation of the associative algebra:
\begin{equation}\label{eq_2.131}
{\rm ad}_{C}(AB)=\{C,AB\}=A\{C,B\}+\{C,A\}B=\left({\rm ad}_{C}A\right)B+A\left({\rm ad}_{C}B\right).
\end{equation}

The both structures may be transported from the function space over $G^{\prime\ast}$ into function space over $G^{\prime}$ by means of the Fourier transform. The pointwise product in $G^{\prime\ast}$ becomes the convolution in $G^{\prime}$. All relationships are preserved. The new Poisson bracket in $G^{\prime}$ is a differentiation of the Abelian convolution.

Let us denote the corresponding Poisson bracket in $G^{\prime}$ by $[,]$. More precisely, if $F$, $G$ are functions on $G^{\prime}$ Fourier-expressed as
\begin{eqnarray}
F(\overline{\omega})&=&\frac{1}{(2\pi\hbar)^{n}}\int \widehat{F}(\underline{\sigma})\exp\left(\frac{i}{\hbar}\
\underline{\sigma}\cdot\overline{\omega}\right)d_{n}\underline{\sigma},
\label{eq_2.132a}\\
G(\overline{\omega})&=&\frac{1}{(2\pi\hbar)^{n}}\int \widehat{G}(\underline{\sigma})\exp\left(\frac{i}{\hbar}\
\underline{\sigma}\cdot\overline{\omega}\right)d_{n}\underline{\sigma},
\label{eq_2.132b}
\end{eqnarray}
then their bracket is defined as
\begin{equation}\label{eq_2.133}
[F,G]\left(\overline{\omega}\right)=
\frac{1}{(2\pi\hbar)^{n}}\int \{\widehat{F},\widehat{G}\}(\underline{\sigma})\exp\left(\frac{i}{\hbar}\
\underline{\sigma}\cdot\overline{\omega}\right)d_{n}\underline{\sigma}.
\end{equation}
One can show that
\begin{equation}\label{eq_2.134}
[F,G]=\frac{1}{i\hbar}\left(\mathcal{A}_{a}F\right)\ast
\left(\omega^{a}G\right)=\frac{1}{i\hbar}\mathcal{A}_{a}\left(F\ast
\omega^{a}G\right).
\end{equation}
Concerning the last formula, let us notice that
\begin{equation}\label{eq_2.135}
\mathcal{A}_{a}\left(f\ast g\right)=\left(\mathcal{A}_{a}f\right)\ast g+
f\ast\left(\mathcal{A}_{a}g\right),
\end{equation}
but it may be also shown that for any $G$
\begin{equation}\label{eq_2.136}
\mathcal{A}_{a}\left(\omega^{a}G\right)=0.
\end{equation}
This explains why only one term appears in the middle expression in (\ref{eq_2.134}). Another, equivalent expression for $[F,G]$ is 
\begin{equation}\label{eq_2.137}
[F,G]=-\frac{1}{i\hbar}\mathcal{A}_{a}\left(\left(\omega^{a}F\right)\ast
G\right)=-\frac{1}{i\hbar}\left(\omega^{a}F\right)\ast
\left(\mathcal{A}_{a}G\right).
\end{equation}
Therefore, the more symmetric formula for $[F,G]$ would be
\begin{eqnarray}
[F,G]&=&\frac{1}{i\hbar}\left(\left(\mathcal{A}_{a}F\right)\ast
\left(\omega^{a}G\right)-\left(\omega^{a}F\right)\ast
\left(\mathcal{A}_{a}G\right)\right)\nonumber\\
&=&\frac{1}{i\hbar}\left(\left(\mathcal{A}_{a}F\right)\ast
\left(\omega^{a}G\right)-\left(\mathcal{A}_{a}G\right)
\ast\left(\omega^{a}F\right)\right).\label{eq_2.138}
\end{eqnarray}

Let us stress here some subtle point concerning the relationship between symbols $k^{a}$, $\omega^{a}$. Roughly speaking, they denote almost the same, however, some delicate difference in their meaning should be noted. In (\ref{eq_2.6}) the canonical coordinates $k^{a}$ are analytically used as coefficients at the basic elements $e_{a}$ of the Lie algebra $G'$. Being used as a parametrization of $G^{\prime}$, they are functions on the group manifold $G$, in general in a local sense. The exponential mapping $\exp$ of $G'$ into $G$ establishes a correspondence between $k^{a}$ and $\omega^{a}$, namely, $\omega^{a}=k^{a} \circ \exp$, when carefully taking domains into account. One must remember however that strictly speaking, $k^{a}$ as functions on $G$ are defined locally and the range of their values is not identical with $\mathbb{R}^{n}$. Unlike this, $\omega^{a}$ are global linear coordinates on the linear space $G'$. Interpretation of functions on $G$ in terms of functions on $G'$ is also local; as a rule, the global identification fails, even because of simple topological reasons. The point is, however, that in the quasiclassical limit these obstacles become inessential. In this limit we deal with "large quantum numbers", i.e., with "quickly oscillating" functions. One performs some truncation or cut-off procedure, namely, the total group algebra over $G$ is replaced by its subalgebra composed of ideals $M(\alpha)$ the generating units $\varepsilon(\alpha)$ of which have the number of nods above some fixed value. The higher is the truncation threshold, the more is the essential behaviour of admissible functions concentrated in a small neighbourhood of the group unity $e$. The admissible functions on $G$ practically vanish far away from $e$, and "do not feel" the topology of $G$. They may be in a good approximation represented by functions on $G'$, thus, on a linear space. More precisely, it is so for functions superposed in a quasiclassical way of the basic quickly oscillating functions $\varepsilon(\alpha)_{ij}$. By that we mean that the combination coefficients $C(\alpha)_{ij}$ are concentrated in a "wide range" of the label $\alpha$ and are "slowly varying" within that range. To be more (even if roughly) rigorous with such statements, one must specify what is meant when we say that the labels $\alpha$, $\beta$ are nearby. Simply we mean then that the numbers of nodes of $\varepsilon(\alpha)$, $\varepsilon(\beta)$ are nearby (roughly speaking, the corresponding quantum numbers are nearby). Functions on $G$ constructed according to such prescription may be reasonably represented by functions on the Lie algebra $G'$. Operations in the group algebra of $G$ may be approximated by certain operations in the group algebra of $G'$, where, just as above, $G'$ is interpreted as an Abelian additive Lie group. Continuous Fourier expansion approximates in a satisfactory way the discrete Peter-Weyl expansion on the compact group $G$. Expanding in the convolution formulas the group multiplication rule in Taylor series and retaining the lowest-order terms, we obtain some asymptotic approximate formulas, namely,
\begin{equation}
F \underset{G}{\ast} H \approx F \underset{G'}{\ast} H + \frac{i\hbar}{2}[F,H], \label{eq_2.139}
\end{equation}
where $[F,H]$ is just (\ref{eq_2.137}), (\ref{eq_2.138}) and the symbols $\underset{G}{\ast}$, $\underset{G^{\prime}}{\ast}$ denote respectively convolutions in the sense of $G$ and $G^{\prime}$ (as an additive group). The use of the same symbols $F$, $H$ on the left and right sides of (\ref{eq_2.139}) is rough, however, the meaning is obvious: just the "identification" in terms of the exponential map. In the lowest order of approximation, the quantum Poisson bracket is expressed as follows:
\begin{equation}\label{eq_2.140}
\{F,H\}_{\rm q}=\frac{1}{i\hbar}\left(F\underset{G}{\ast}H-
H\underset{G}{\ast}F\right)\approx[F,H].
\end{equation}
Obviously, (\ref{eq_2.139}) and (\ref{eq_2.140}) is a counterpart of the well-known quasiclassical expansion of star products, first of all, the Weyl-Moyal product.

Some more details will be presented when discussing the physically important special case $G={\rm SU}(2)$ or $G={\rm SO}(3,\mathbb{R})$, i.e., quantum description of angular momentum.

\section*{Acknowledgements}

This paper partially contains results obtained within the framework of the research project 501 018 32/1992 financed from the Scientific Research Support Fund in 2007-2010. The authors are greatly indebted to the Ministry of Science and Higher Education for this financial support. The support within the framework of Institute internal programme 203 is also greatly acknowledged.


\begin{thebibliography}{99}

\bibitem{12}
W. Ambrose, {\it Amer. Math. Soc.}, {\bf 57}, 364, 1945.

\bibitem{5}
L. E. Ballentine, {\it Quantum Mechanics: A Modern Development}, World Scientific Publishing Co. Ltd., Singapore-New Jersey-London-Hong Kong, 1998.

\bibitem{Godl}
P. Godlewski, {\it Quantization of Anisotropic Rigid Body}, International Journal of Theoretical Physics, {\bf 42}, no. 12, 2863--2875, 2003.

\bibitem{21}
A. A. Kirillov, {\it \'{E}l\'{e}ments de la Th\'{e}orie des Repr\'{e}sentations}, \'{E}ditions MIR, Moscow, 1974.

\bibitem{22}
A. A. Kirillov, {\it Merits and Demerits of the Orbit Method}, Bull. Amer. Math. Soc., {\bf 36}, 433--488, 1999.

\bibitem{20}
A. A. Kirillov, {\it Lectures on the Orbit Method}, Graduate Studies in Mathematics, {\bf 64}, American Mathematical Society, Providence, RI, 2004.

\bibitem{3}
L. D. Landau, E. M. Lifshitz, {\it Course of Theoretical Physics. Vol. III. Quantum Mechanics}, Pergamon Press, London, 1958.

\bibitem{2}
L. H. Loomis, {\it An Introduction to Abstract Harmonic Analysis}, D. Van Nostrand Company, Inc., Princeton-New Jersey-Toronto-London-New York, 1953.

\bibitem{1a}
G. W. Mackey, {\it The Mathematical Foundations of Quantum Mechanics}, Benjamin, New York, 1963.

\bibitem{11}
K. Maurin, {\it General Eigenfunction Expansions and Unitary Representations of Topological Groups}, PWN --- Polish Scientific Publishers, Warsaw, 1968.

\bibitem{13}
K. Maurin, {\it Methods of Hilbert Spaces}, PWN --- Polish Scientific Publishers, Warsaw, 1972.

\bibitem{14}
K. Maurin, {\it Analysis. Part I--III}, D. Reidel-PWN, Dordrecht-Warszawa, 1980.

\bibitem{10}
L. Pontryagin, {\it Topological Groups}, Princeton University Press, Princeton, New Jersey, 1956.

\bibitem{4}
M. E. Rose, {\it Elementary Theory of Angular Momentum}, Dover Publications, 1965.

\bibitem{Rudin}
W. Rudin, {\it Fourier Analysis on Groups}, Interscience Publ., New York-London, 1962.

\bibitem{8}
F. E. Schroeck, Jr., {\it Quantum Mechanics on Phase Space}, Kluwer Academic Publishers, Dordrecht-Boston-London, 1996.

\bibitem{17}
J. J. S\l awianowski, {\it Abelian Groups and the Weyl Approach to Kinematics. Non-Local Function Algebras}, Reports on Mathematical Physics, {\bf 5}, no. 3, 295--319, 1974.

\bibitem{18}
J. J. S\l awianowski, {\it Geometry of Phase Spaces}, PWN --- Polish Scientific Publishers, Warsaw; John Wiley \& Sons, Chichester-New York-Brisbane-Toronto-Singapore, 1991.

\bibitem{16}
J. J. S\l awianowski, V. Kovalchuk, {\it Schr\"{o}dinger and Related Equations as Hamiltonian Systems, Manifolds of Second-Order Tensors and New Ideas of Nonlinearity in Quantum Mechanics}, Reports on Mathematical Physics, {\bf 65}, no. 1, 29--76, 2010; arXiv:0812.5055.

\bibitem{19}
J. J. S\l awianowski, V. Kovalchuk, B. Go\l ubowska, A. Martens, E. E. Ro\.{z}ko, {\it Quantized Excitations of Internal Affine Modes and Their Influence on Raman Spectra}, Acta Physica Polonica B, {\bf 41}, no. 1, 165--218, 2010; arXiv:0901.0243.

\bibitem{1} J. J. S\l awianowski, V. Kovalchuk, A. Martens, B. Go\l ubowska, E. E. Ro\.{z}ko, {\it Quasiclassical and Quantum Systems of Angular Momentum. Part I. Group Algebras as a Framework for Quantum-Mechanical Models with Symmetries}, arXiv:1007.4121.

\bibitem{9}
W. M. Tulczyjew, {\it The Theory of Systems with Internal Degrees of Freedom}, Lecture Notes, Department of Physics, Lehigh University, Bethlehem, Pennsylvania, 1964.

\bibitem{6}
H. Weyl, {\it The Theory of Groups and Quantum Mechanics}, Dover, New York, 1950.

\bibitem{15}
H. Weyl, {\it Symmetry}, Princeton University Press, Princeton, New Jersey, 1952.

\bibitem{7}
E. P. Wigner, {\it Gruppentheorie und Ihre Anwendungen auf die Quantenmechanik der Atomspektren}, Vieweg Verlag, Braunschweig, 1931. English Translation by J. J. Griffin, {\it Group Theory and its Application to the Quantum Mechanics of Atomic Spectra}, Academic Press, New York, 1959.

\end{thebibliography}
\end{document}